\documentclass{aa}
\usepackage{graphicx}

\usepackage{color}
\usepackage{txfonts}
\usepackage{comment}
\usepackage{soul}

\usepackage{amssymb}
\usepackage{mathrsfs}
\usepackage{minted}

\usepackage[colorlinks=true, allcolors=blue]{hyperref}

\usepackage{amsmath,amsfonts}
\usepackage{xcolor}
\usepackage{enumitem}
\usepackage{booktabs}
\usepackage[normalem]{ulem}

\usepackage{threeparttable}
\usepackage[table]{colortbl}
\usepackage{soul}
\usepackage{subcaption}

\begin{document}

   \title{Properties of stellar disks and spheroids in simulated galaxies}

   \author{Valeria A. Cristiani
          \inst{1,2,3}\thanks{e-mail: valeria.cristiani@unc.edu.ar},
          Julio F. Navarro\inst{4,5}
          \and
          Mario G. Abadi\inst{1,2}
          }

   \institute{CONICET. Instituto de Astronomía Teórica y Experimental. Córdoba, Argentina.
           \and
             Universidad Nacional de Córdoba. Observatorio Astronómico de Córdoba. Laprida 854, X5000BGR, Córdoba, Argentina.
         \and
             Facultad de Matem\'atica, Astronom\'{\i}a, F\'{\i}sica y Computaci\'{\o}n, Universidad Nacional de C\'ordoba (FaMAF--UNC). Bvd. Medina Allende s/n, Ciudad Universitaria, X5000HUA, C\'ordoba, Argentina.
        \and
            Department of Physics and Astronomy, University of Victoria, Victoria, Canada
        \and
            KU-KIST Green School, Korea University
             }

    \titlerunning{}
    \authorrunning{Cristiani et al.} 
   \date{Received 09 April 2026; accepted 07 November 2026}

\abstract{The optical morphology of a galaxy is determined by the relative importance of their disk and spheroidal components. Although the scaling laws of disk-dominated or spheroid-dominated galaxies have been studied in some detail, less attention has been paid to the properties of each component in galaxies where both coexist. We use a dynamical decomposition of 3452 galaxies from the TNG100 simulations with stellar masses in the range $10 \leq \log_\mathrm{10}(M_*/M_\odot) \leq 12$ to study how, at given mass, the properties of a spheroid are altered by the presence of a disk, and vice versa. We define spheroids as the non-rotating component made up of all stars in retrograde motion, plus an equal number of co-rotating stars mirrored in angular momentum and energy. Disks simply consist of the remainder of (co-rotating) stars. We find that disk properties (rotation speed, size, formation time) depend sensitively on the mass of the spheroidal component and therefore of the total mass of the galaxy in which both reside. At given mass, disks rotate faster, are more extended, and form their stars earlier as the mass of the spheroidal component increases. On the other hand, at fixed mass, the properties of spheroids are less sensitive to the presence of a disk. Roughly speaking, the classical bulges of simulated disk galaxies closely resemble ellipticals of similar stellar mass. The apparent structural invariance of spheroids to the presence of a disk is somewhat surprising, but it likely results because disks form out of high angular momentum material which surrounds the spheroid and thus have less dynamical influence on a spheroid's structure.}


   \maketitle
\nolinenumbers

\section{Introduction}

Galaxies are complex stellar systems whose present-day structure reflects the cumulative effects of hierarchical assembly, gas accretion, star formation, and dynamical evolution. In the context of the $\Lambda$CDM paradigm, galaxies form due to the gravitational collapse of gas in an evolving hierarchy of dark matter halos. As gas cools, it falls to the center of the halos, where it can turn into stars and form a galaxy \citep{White-1978, Fall-1980}. Galaxies thus evolve through physical processes that involve a wide range of temporal and spatial scales,  baryon accretion, radiative cooling, star formation, and feedback from supernovae and active galactic nuclei (AGN), as well as galaxy mergers of smaller systems \citep[see, for example][]{Naab-2017}. The combination of these astrophysical processes, together with the influence of the environment, shapes galaxies and gives rise to the observed diversity of galaxy morphology and internal structure \citep[e.g.][]{Kartaltepe-2015, Willett-2017}.

The complexity of galaxy formation is reflected in the presence of different structural components, such as thin and thick disks, spiral arms, bars, bulges, and stellar halos \citep{Buta-2013}. From a dynamical standpoint, and as a first-order approximation, these structural properties may be summarized as the superposition of two dominant stellar components: a rotationally-supported disk and a velocity dispersion-supported spheroidal component. The relative importance of these components underlies the classical morphological sequence, from disk-dominated late-type galaxies to spheroid-dominated early-type systems \citep{Hubble-1926}. While this conceptual framework is well established, the extent to which disks and spheroids influence each other’s structural and evolutionary properties within the same galaxy is less well understood.

Observational studies have traditionally characterized stellar components through photometric decompositions of surface brightness profiles, often assuming exponential disks and Sérsic spheroids \citep{Freeman-1970, de_Vaucouleurs-1959, Sersic-1963}. Large surveys have enabled statistical analyses of component masses, sizes, and scaling relations across morphological types \citep[e.g.][]{Shen-2003, Lange-2016, Moffett-2016, Casura-2022}. However, photometric decompositions are affected by projection effects, degeneracies between components, and uncertainties in the conversion from light to stellar mass. Kinematic decompositions based on integral-field spectroscopy provide additional constraints but remain limited to relatively small samples and specific environments \citep[e.g.][]{Zhu-2018, Tabor-2019}.

Cosmological hydrodynamical simulations offer a complementary route to study galaxy structure by providing direct access to the phase-space information of stellar particles. State-of-the-art simulations such as IllustrisTNG reproduce many observed galaxy properties, including stellar mass functions, sizes, and kinematic scaling relations, making them suitable tools for investigating the internal structure of galaxies \citep{Pillepich-2018, Marinacci-2018, Nelson-2018, Nelson-2019, Springel-2018}. In this framework, stellar components can be identified through dynamical decomposition methods that separate rotationally supported and dispersion-dominated stellar populations based on their orbital properties \citep[e.g.][]{Abadi-2003, Scannapieco-2009, Tissera-2012, Obreja-2018, Du-2019, Zana-2022}.

Previous simulation-based studies have used such techniques to quantify the mass fractions of disks and spheroids, explore their scaling relations, and investigate their dependence on halo mass and environment \citep[e.g.][]{Marinacci-2014, Gargiulo-2019, Du-2020, Kumar-2021}. This body of work has firmly established that galaxy morphology correlates with stellar mass and halo mass, and that disks and spheroids exhibit distinct kinematic and structural properties \citep[e.g.,][]{Ferrero-2021}. However, most existing analyses treat disks and spheroids either as independent components or focus on global galaxy properties as a function of morphology.

This body of work has found that stellar disks are more extended and their stellar populations are bluer, younger, and more metal-rich compared to bulges or stellar halos \citep{Tissera-2012, Domenech-2012, Gargiulo-2019, Obreja-2019, Tacchella-2019, Jagvaral-2022}. In addition, disk-dominated or late-type galaxies tend to have higher star formation activity in comparison with galaxies where the spheroidal component dominates \citep{Tacchella-2019, Du-2020, Du-2021, Ferrero-2021, Jagvaral-2022, Rodriguez-Gomez-2022}.

What remains less explored is whether the intrinsic properties of dynamically identified stellar components depend on the relative prominence of the other component in the same galaxy. In particular, it is not yet clear whether disks of fixed stellar mass retain the same structural and evolutionary properties across galaxies with different spheroid mass fractions, or whether spheroids are sensitive to the presence and mass of an accompanying disk. Addressing these issues is essential for assessing whether disks and spheroids evolve largely independently or whether their properties are coupled through shared assembly histories and dynamical interactions.

In this work, we use a large sample of central galaxies drawn from the TNG100 simulation to investigate this question. We apply a dynamical decomposition method to identify stellar disks and spheroids and analyze their properties separately. Our analysis focuses on how characteristic sizes, kinematics, and formation times of each component depend not only on their own stellar mass, but also on the relative importance of the other component. Our results provide new constraints on the degree of coupling between disks and spheroids and offer insight into the physical origin of galaxy morphology.

This paper is organized as follows. Section 2 describes the simulations and the galaxy sample. Section 3 presents the dynamical decomposition method. Sections 4 and 5 analyze global galaxy properties and the intrinsic properties of stellar disks and spheroids, respectively. Section 6 summarizes our conclusions.

\section{Numerical methods}
\label{Sec:Data}

\subsection{The IllustrisTNG run}

We use data from the TNG100 run of the IllustrisTNG project. IllustrisTNG is a suite of state-of-the-art cosmological galaxy formation simulations available publicly. We describe them briefly below, and refer the interested reader to the original papers for more details \citep{Pillepich-2018, Naiman-2018, Nelson-2018, Nelson-2019, Marinacci-2018, Springel-2018}.

Like its predecessor series, Illustris \citep{Vogelsberger-2014}, the IllustrisTNG simulations (TNG, for short) employ the {\sc AREPO} code \citep{Springel-2010}. The TNG simulations evolve different cosmological volumes, from $z=127$ to $z=0$ using an adaptive hydro mesh. The adopted cosmological parameters correspond to the \citet{Planck-2016} results. These simulations include models of astrophysical processes, such as gas cooling, star formation, stellar evolution, and feedback from AGN and supernova. For more details about the models and how they are implemented, see \citet{Pillepich-2018}.

The TNG100 run corresponds to a $110.7$-Mpc box containing $1820^{3}$ dark matter particles that interact with a gravitational softening $\epsilon = 0.74$ kpc. The mass resolution is $m_\mathrm{dm} = 7.5 \times 10^6\ M_\odot$ for dark matter particles and $m_\mathrm{gas} = 1.4 \times 10^6\ M_\odot$ for gas \citep{Nelson-2019}. The halo catalog is built up from a Friends-of-Friends groupfinder output \citep[][FoF]{Davis-1985} and a subhalo catalog constructed using SUBFIND \citep{Springel-2001, Dolag-2009}.

\subsection{The simulated galaxy sample}

We define a galaxy as all the star, dark matter, and gas particles that are gravitationally bound according to SUBFIND and located within three times the stellar half-mass radius \citep{Du-2020}. These particles are used to define the $z$-axis of the coordinate system, chosen to coincide with the direction of their total angular momentum. These particles are then used to calculate all galactic properties and to perform the dynamical decomposition.

We selected central galaxies (i.e., the main subhalos of a given FoF group) from the subhalo catalog at redshift $z=0$ \citep{Nelson-2019}, with stellar masses in the range $10~\leq~\log_\mathrm{10}(M_*/M_\odot)~\leq~12$. The lower limit has been chosen so that galaxies have at least $\sim 7\,000$ stellar particles. This sample comprises 3566 galaxies. 

In many cases, we noticed that galaxies show clear signs of perturbations, mainly due to recent accretion events and mergers. We quantify how disturbed the systems are by using the $m=0$ and $m=1$ Fourier modes of the galaxy stellar mass distribution. Defining
\begin{equation}
    a_m = \sum^{N_\mathrm{part}}_{i=1} m_i\,cos(m\,\phi_i)
\end{equation}
and
\begin{equation}
    b_m = \sum^{N_\mathrm{part}}_{i=1} m_i\,sin(m\,\phi_i),
\end{equation}
we use $A_1 = \sqrt{a_1^2 + b_1^2}/a_0$ as a measure of the asymmetry of a galaxy projected onto the $x$-$y$ plane. Here $N_\mathrm{part}$ is the number of galaxy star particles, $m_i$ the mass of the $i-th$ particle, and $\phi_i$ its azimuthal angle. We discard galaxies with $A_1 > 0.1$, which, upon visual inspection, removes the most obviously perturbed galaxies. With this restriction, the final sample has a total of 3452 galaxies.

\section{Dynamical decomposition}
\label{Sec:Method}

\subsection{The method}

Reliably identifying the dynamical components of galaxies is not a trivial task. In simulated galaxies, where masses, positions, and velocities are available, integrals of motion can be used to assign stars to different stellar components. In this work, we apply the method introduced by \citet{Cristiani-2024}, which is an improved version of the original implementation developed by \citet{Abadi-2003}. The particular method used here to dynamically identify the stellar disk and spheroidal components was compared with five other methods widely used in the literature \citep[e.g.,][]{Abadi-2003, Domenech-2012, Vogelsberger-2014, Obreja-2018, Du-2019}, and shown to recover the intrinsic galaxy characteristics (e.g. scale lengths, density profiles, velocity profiles) substantially better. We briefly explain below how the method works; see \citet{Cristiani-2024} for a more detailed description.

The method assigns all stellar particles of a galaxy to either a spheroidal or a disk component. The assignment is performed using two quantities: the circularity parameter ($\epsilon$) and the binding energy ($E$). The circularity parameter is defined as
\begin{equation}
    \epsilon = J_\mathrm{z} / J_\mathrm{circ}(E)
\end{equation}
where $J_\mathrm{z}$ is the $z$-component of the angular momentum of each particle and $J_\mathrm{circ}(E)$ is the angular momentum of a star particle with the same binding energy in a co-rotating circular orbit.

The circularity parameter describes the orbit of each particle. For example, $\epsilon \sim 1$ corresponds to a co-rotating circular orbit, $\epsilon \sim -1$ to a counter-rotating circular orbit, and $\epsilon \sim 0$ to an orbit with no net sense of rotation around the $z$ axis. The binding energy indicates whether a particle is bound ($E < 0$) or unbound ($E \geq 0$) to the galaxy.

The method assumes that the disk component does not contain counter-rotating particles. Consequently, all particles with $\epsilon \leq 0$ are assigned to the spheroidal component. The method also assumes that the spheroidal component does not rotate. Therefore, the method assigns to the spheroid also co-rotating particles that match the binding energy distribution of the counter-rotating component, canceling out the net $z$ angular momentum of the spheroid. The disk is made out of all remaining co-rotating particles.

\subsection{Disk vs spheroid mass}

\begin{figure}
    \centering
    \includegraphics[width=\columnwidth]{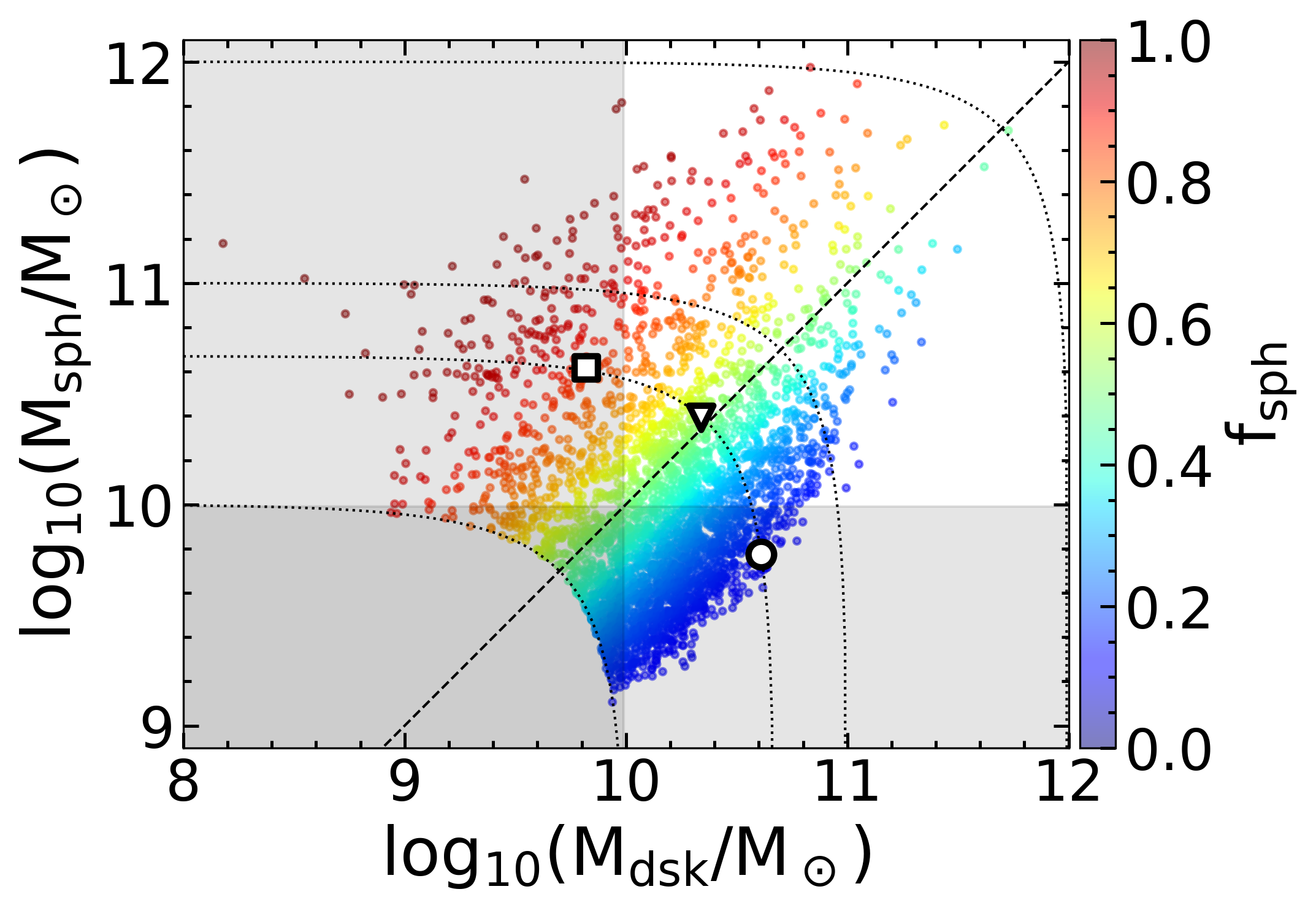}
    \caption{Spheroid stellar mass versus disk stellar mass for each galaxy. Each point is colored by the spheroidal mass fraction of the galaxy according to the colorbar on the right. The dashed line corresponds to the 1:1 relation. The dotted lines correspond to fixed galaxy stellar mass; from bottom to top, $M_* = 10^{10}M_\odot, \ 10^{10.67}M_\odot, \ 10^{11}M_\odot,\ 10^{12}M_\odot$. The white circle, triangle, and square symbols indicate the three example galaxies shown in Fig.~\ref{Fig:gal_example}, with $f_\mathrm{sph}= 0.13,\ 0.53,\ 0.86$, respectively. The shaded vertical areas indicate galaxies with either $M_{\rm dsk}$ or $M_{\rm sph}<10^{10}\, M_\odot$.}
    \label{Fig:Msph_Mdsk}
\end{figure}

\begin{figure*}
    \centering
    \includegraphics[width=\textwidth]{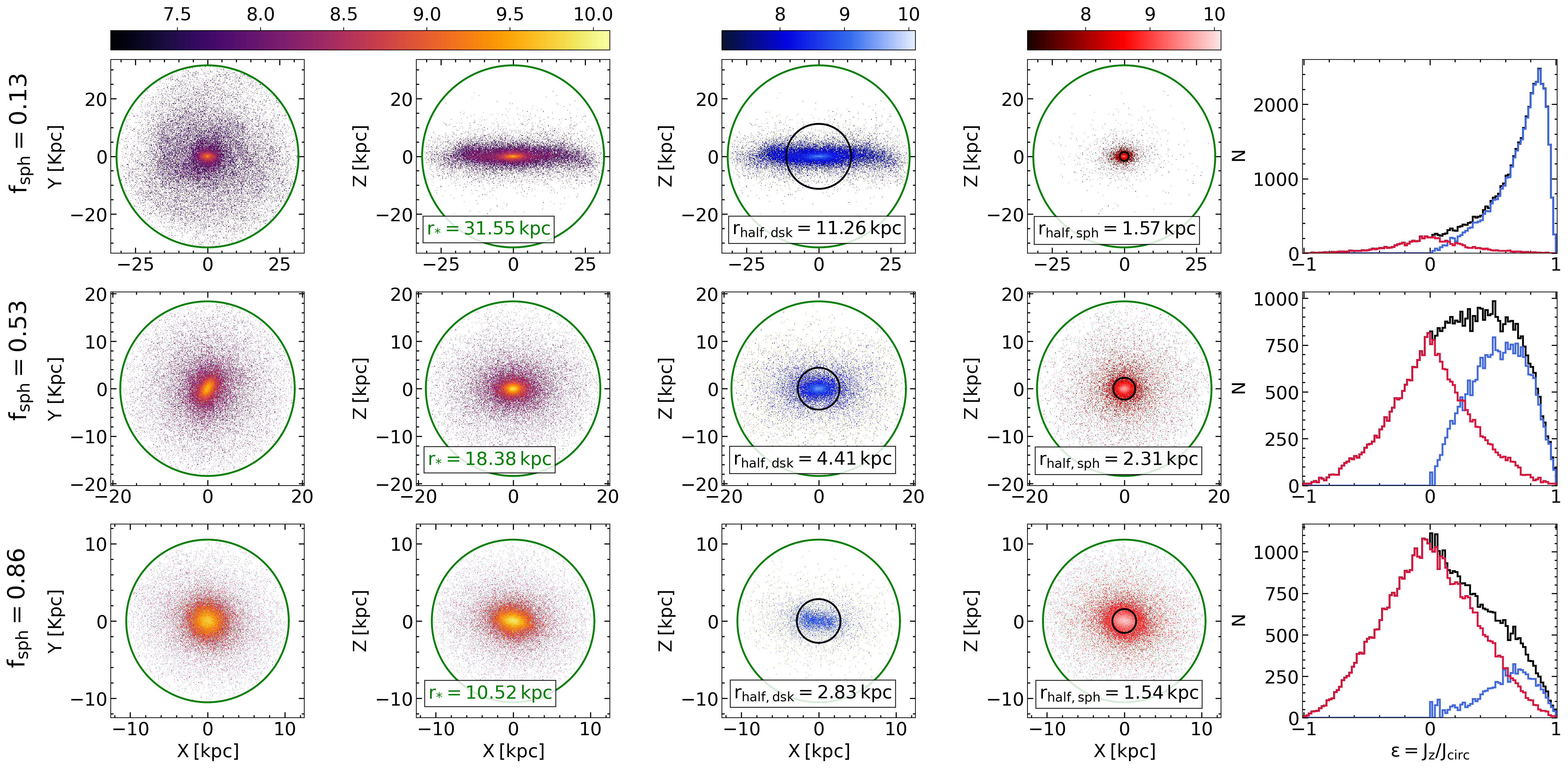}
    \caption{Dynamical decomposition of three TNG100 galaxies with approximately the same $\log_\mathrm{10}(M_*)~\sim~10.67$, but with different spheroidal mass fractions. These galaxies have $f_\mathrm{sph}=0.13$ (first row), $f_\mathrm{sph}=0.53$ (second row), and $f_\mathrm{sph}=0.86$ (third row). For each galaxy, the first two columns show the stellar surface density in a face-on view and an edge-on view, respectively. The third and fourth columns show the edge-on view of the stellar surface density of the disk and spheroid components, respectively. The fifth column shows the circularity parameter histogram of the stellar particles used for dynamic decomposition (black) and the resulting contributions of the disk (blue) and spheroidal (red) components. In all cases, the color bar bars indicate stellar surface densities in units of $\log_\mathrm{10}(M_\odot$ kpc$^{-2})$; the green circle shows the galaxy radius, defined as three times the stellar half-mass radius according to SUBFIND. Black circles indicate the stellar half mass radius of each component.}
    \label{Fig:gal_example}
\end{figure*}

The total stellar mass of the galaxy is
\begin{equation}\label{Eq:masa}
    M_* = M_\mathrm{sph} + M_\mathrm{dsk}
\end{equation}
where $M_\mathrm{sph}$ and $M_\mathrm{dsk}$ are the stellar masses of the spheroidal and disk components, respectively. In what follows, we shall use the spheroid mass fraction ($f_\mathrm{sph}=M_\mathrm{sph}/M_*$) as a parameter to quantify the morphology of a galaxy. Figure~\ref{Fig:Msph_Mdsk} shows $M_\mathrm{sph}$ versus $M_\mathrm{dsk}$ for each simulated galaxy in our sample, colored by $f_\mathrm{sph}$. The three dotted lines indicate galaxies of fixed total stellar mass: from bottom to top, $M_*=10^{10}\,M_{\odot},\ 10^{10.67}\,M_{\odot},\ 10^{11}\,M_{\odot},\ 10^{12}\,M_{\odot}$. Note that for a fixed disk  mass, increasing $f_{\rm sph}$ increases the total stellar mass of the galaxy. 

The lack of points in the lower left corner of Fig.~\ref{Fig:Msph_Mdsk} is due to the lower limit on $M_*>10^{10}\, M_\odot$ imposed on our sample. This figure also makes clear that the lower limit in $M_*$ imposes restrictions on the mass range of disks (or spheroids) we can consider in our analysis. For example, in our sample, $10^9\, M_\odot$ disks can only exist in spheroid-dominated systems (i.e., $f_{\rm sph}>0.9$), precluding a fair analysis of disks of that mass in the presence of a spheroidal component. A similar comment applies to low-mass spheroids. We shall therefore restrict our analysis in some cases to disks with $M_{\rm dsk}>10^{10}\, M_\odot$ and, analogously, to spheroids with $M_{\rm sph}>10^{10}\, M_\odot$, as highlighted by the regions not shaded in gray in Fig.~\ref{Fig:Msph_Mdsk}.

\subsection{Dynamical decomposition example}

Fig.~\ref{Fig:gal_example} illustrates the result of our dynamical decomposition algorithm applied to three TNG100 galaxies of similar stellar mass, $\log_\mathrm{10}(M_*) \sim 10.67$, but different values of spheroid mass. The first row shows a disk-dominated galaxy with $f_\mathrm{sph} = 0.13$, the second row an intermediate galaxy with $f_\mathrm{sph} = 0.53$, and the third row a spheroid-dominated galaxy with $f_\mathrm{sph} = 0.86$. 

The first two columns in Fig.~\ref{Fig:gal_example} show the face-on and the edge-on projected stellar mass density distribution of each galaxy, respectively. The third and fourth columns correspond to the edge-on projection of their disk and spheroid components, respectively. Finally, the fifth column shows the circularity parameter distribution of all stellar particles (black), as well as the contributions from the disk (blue), and spheroid (red). Note that the circularity distribution of the spheroidal component is always symmetric about $\epsilon=0$, by construction. The method also makes sure that the energy distribution of co- and counter-rotating spheroid particles is nearly identical. 

The disk-dominated galaxy shows a peak at $\epsilon\sim 1$, which indicates that most of its stars are in co-rotating orbits (see upper row, rightmost column in Fig.~\ref{Fig:gal_example}). As expected, the spheroid-dominated galaxy shows a nearly symmetrical $\epsilon$ distribution about $\epsilon\sim 0$, indicative of a small net angular momentum (see lower row, rightmost column in Fig.~\ref{Fig:gal_example}).

\section{Simulated galaxy properties} \label{Sec:properties of galaxies with fsph}

\subsection{The galaxy mass-halo mass relation}
\label{Subsec:galaxy stellar mass-halo mass relation}

\begin{figure}
    \centering
    \includegraphics[width=\columnwidth]{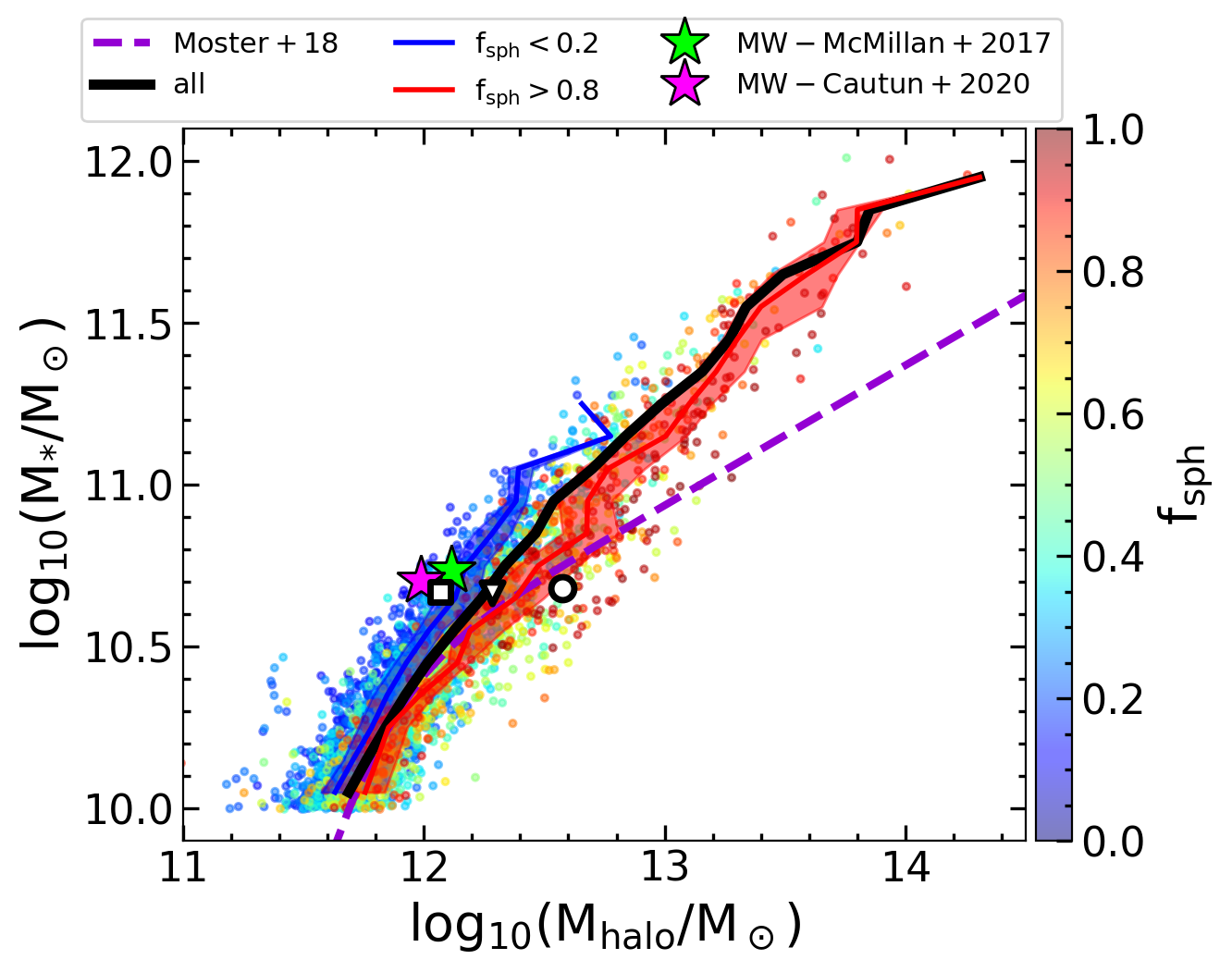}
    \caption{Galaxy stellar mass as a function of halo mass. Each galaxy is colored according to the colorbar on the right. The black line shows the median halo mass calculated in bins of $\log_\mathrm{10}(M_{*}/M_\odot)$. The red (blue) solid line shows the median calculated for those galaxies with $f_\mathrm{sph}~>~0.8$ ($f_\mathrm{sph}~<~0.2$) and the shaded area encloses the 25th and 75th percentiles. The white symbols indicate the three example galaxies shown in Fig.~\ref{Fig:gal_example}, with $f_\mathrm{sph}= 0.13,\ 0.53,\ 0.86$, respectively. For reference, the violet dashed line corresponds to the result of the abundance-matching model of \citet{Moster-2018}, while the green star corresponds to the Milky Way model of \citet{McMillan-2017} and the pink star corresponds to the Milky Way model of \citet{Cautun-2020}.}
    \label{Fig:abundance}
\end{figure}

Figure~\ref{Fig:abundance} shows the galaxy stellar mass as a function of halo virial\footnote{We measure virial quantities inside a (virial) radius which encloses a mean density equal to 200 times the critical density for closure.} mass, where each galaxy is colored according to its corresponding spheroidal mass fraction $f_\mathrm{sph}$. The solid black line tracks the median $M_*$ of the complete sample, whereas the violet dashed line shows the abundance-matching model of \citet{Moster-2018}, for reference. 

Note that, at fixed halo mass, galaxies with higher spheroidal mass fractions tend to have a lower total stellar mass (i.e., lower galaxy formation "efficiency"). Conversely, the higher the disk fraction, the higher the total stellar mass of a galaxy. This is highlighted when we compare the median trends of spheroid-dominated galaxies (defined hereafter as $f_\mathrm{sph} > 0.8$, solid red line) and of disk-dominated galaxies (defined, analogously, as $f_\mathrm{sph} < 0.2$, solid blue line). In simple terms, this difference arises because disk-dominated galaxies form stars over a longer period of time (indeed, they are still forming stars at present), whereas spheroid-dominated galaxies generally stopped forming stars at early times.

An alternative interpretation arises when we consider galaxies at fixed stellar mass. In this interpretation, the higher the spheroidal mass fraction, the larger the virial mass of the surrounding halo. This is illustrated by the three galaxies of similar $M_*$ but different morphologies selected to illustrate the dynamical decomposition algorithm in Fig.~\ref{Fig:gal_example}. 

The disk-dominated galaxy ($f_\mathrm{sph} = 0.13$) in Fig.~\ref{Fig:abundance} is shown with a white circle, the spheroid-dominated galaxy ($f_\mathrm{sph} = 0.86$) with a white square, and the intermediate galaxy ($f_\mathrm{sph} = 0.53$) with a white triangle, respectively. We also show the location of the Milky Way, according to the models of \citet{McMillan-2017} (green star) and \citet{Cautun-2020} (pink star). The relatively low halo mass of the Milky Way (for its $M_*$) seems consistent with the trend seen in Fig.~\ref{Fig:abundance} for TNG100: high galaxy formation efficiencies are the norm for disk-dominated galaxies. 

\subsection{Spheroid and disk stellar mass function}

\begin{figure}
    \centering
    \includegraphics[width=\columnwidth]{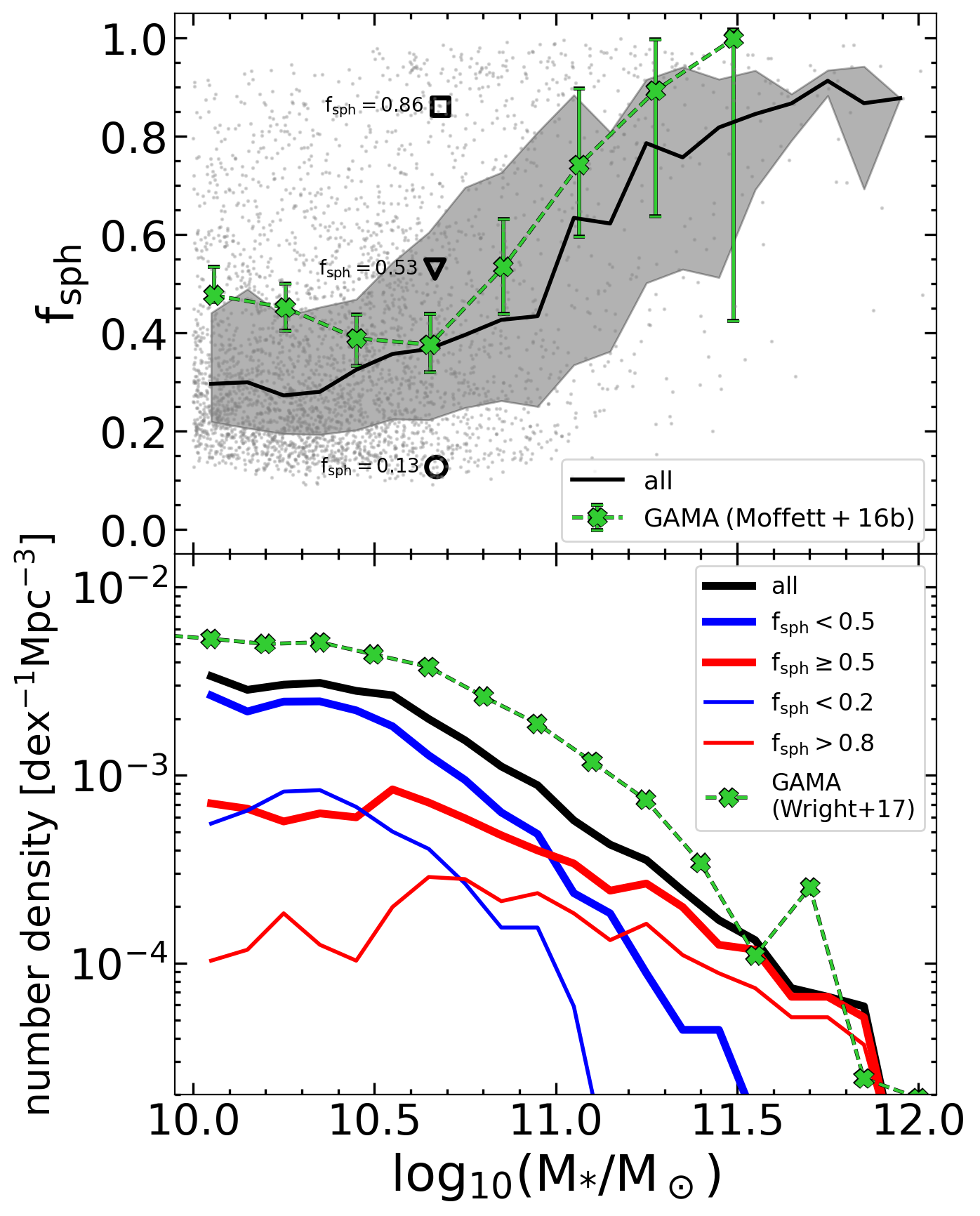}
    \caption{{\it Upper panel}: Spheroid mass fraction as a function of galaxy stellar mass. The gray dots are TNG100 galaxies, the solid black line shows the median trend calculated in bins of the $\log_\mathrm{10}(M_{*}/M_\odot)$, and the gray shading encloses the 25th and 75th percentiles. The dashed green line with crosses shows the spheroid mass fractions estimated for galaxies in the GAMA catalog \citep{Moffett-2016}. White symbols highlight the example galaxies shown in Fig.~\ref{Fig:gal_example}, with $f_\mathrm{sph}= 0.13,\ 0.53,\ 0.86$. {\it Lower panel}: Galaxy stellar mass function. The black line shows the resulting distribution of our TNG100 galaxy sample, which includes only centrals (i.e., no satellites). The red (blue) thick line shows the mass function of galaxies whose spheroid mass fraction is $f_\mathrm{sph}~\geq~0.5$ ($f_\mathrm{sph}~<~0.5$), while the red (blue) thin line corresponds to the distribution of those galaxies whose spheroid mass fraction is $f_\mathrm{sph}~>~0.8$ ($f_\mathrm{sph}~<~0.2$). The GAMA galaxy stellar mass function is shown by the dashed green line with crosses \citep{Wright-2017}.}
    \label{Fig:f_sph}
\end{figure}

The total fraction of stars in disks or spheroids, both as a function of $M_*$ and across the whole galaxy population, is an important metric that constrains models of galaxy formation and evolution. The solid black line in the upper panel of Fig.~\ref{Fig:f_sph} shows the median of the spheroid stellar mass fraction ($f_\mathrm{sph}$) as a function of $M_*$. This shows a gradual increase of spheroid importance with increasing $M_*$: few $10^{10}\, M_\odot$ galaxies are spheroid dominated, but most $10^{11.5}\, M_\odot$ galaxies are. 

This is consistent with the results reported by \citet{Du-2020} for TNG100, although our decomposition method differs from theirs. The trend is also consistent with the observational trend reported by \citet{Moffett-2016} using galaxies from the Galaxy and Mass Assembly (GAMA) survey (see dashed green line with crosses in the upper panel of Fig.~\ref{Fig:f_sph}), although their results, which are based on photometric, rather than dynamical, decomposition show a slightly higher spheroid prevalence at low masses.

The lower panel of Fig.~\ref{Fig:f_sph} shows the TNG100 galaxy stellar mass function (black line), and for reference, the observed galaxy stellar mass function as reported by \citet{Wright-2017} (green crosses). The offset in normalization arises mainly because we are only considering a restricted sample of TNG100 galaxies (i.e., centrals only). As expected from the discussion in the paragraph above, spheroids dominate the high-mass end (red curves), whereas disks prevail at the low-mass end. Splitting the sample at $f_{\rm sph}=0.5$, we see that the contributions of the two populations cross at roughly $M_*\sim 1.25\times 10^{11}\, M_\odot$ (see the thick blue and red curves in Fig.~\ref{Fig:f_sph}).

Adding up the total stellar mass of spheroidal and disk components separately over the whole galaxy sample (i.e., $M_*>10^{10}\, M_\odot$), we find that $46\%$ of all stars are assigned to disks while the remaining $54\%$ reside in spheroidal components. These results are also in agreement with earlier work; for instance, \citet{Schechter-1987} found approximately the same amount of mass in bulge and disk components, based on photometric decomposition of a sample of galaxies in a set of random fields. \citet{Kelvin-2014} and \citet{Moffett-2016a,Moffett-2016} also report similar results for the GAMA survey.

\begin{figure*}
    \centering
    \includegraphics[width=\textwidth]{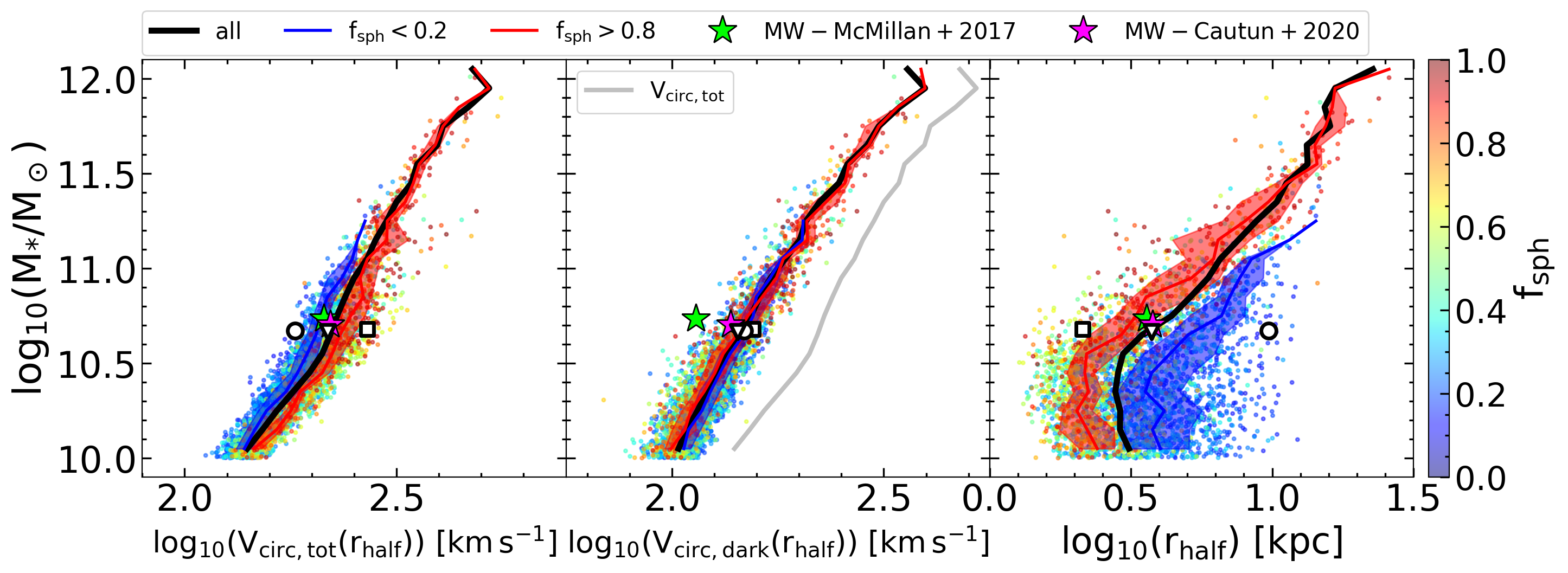}
    \caption{Scaling relations of TNG-100 galaxies. {\it Left panel}: Galaxy stellar mass versus circular velocity measured at the stellar half-mass radius of the galaxy. {\it Center panel}: Galaxy stellar mass versus the contribution of the dark matter to the circular velocity at the stellar half-mass radius of the galaxy. {\it Right panel}: Galaxy stellar mass versus the stellar half-mass radius of the galaxy. The colorbar encodes the spheroidal mass fraction of a galaxy. Black lines track the median trends computed in $\log_\mathrm{10}(M_{*}/M_\odot)$ bins. Red (blue) solid lines correspond to the median trends calculated using only galaxies with $f_\mathrm{sph}~>~0.8$ ($f_\mathrm{sph}~<~0.2$) and the shaded area encloses the 25th and 75th percentiles. The white symbols highlight the three example galaxies shown in Fig.~\ref{Fig:gal_example}, with $f_\mathrm{sph}= 0.13,\ 0.53,\ 0.86$, respectively. For reference, the green and pink stars indicate the Milky Way, according to the \citet{McMillan-2017} and \citet{Cautun-2020} models, respectively. The gray solid line in the central panel reproduces the median trend of the circular velocity shown in the left panel.}
    \label{Fig:vcirc_sgm_re}
\end{figure*}

\subsection{Circular velocity and size}

The morphology of a simulated galaxy (as measured by $f_{\rm sph}$) also affects the relation between stellar mass and the characteristic circular velocity (the ``Tully-Fisher relation"). This is shown in the left panel of Fig.~\ref{Fig:vcirc_sgm_re}, where the relation between $M_*$ and $V_{\rm circ,tot}(r_{\rm half})$, the circular velocity measured at the stellar half-mass radius of the galaxy, is colored by $f_{\rm sph}$. At fixed velocity, disk-dominated galaxies are systematically more massive than spheroid-dominated ones.

This is actually not due to the offset in the galaxy mass-halo mass relation noted when discussing Fig.~\ref{Fig:abundance}. Indeed, as the middle panel of Fig.~\ref{Fig:vcirc_sgm_re} shows, both disk- and spheroid-dominated systems show the same relation with circular velocity when only the dark matter contribution to the circular velocity at the stellar half-mass radius of the galaxy $V_{\rm circ,dark}(r_{\rm half})$ is considered. This lack of difference in dark matter mass between galaxies of the same stellar mass but different morphologies indicates that the offset in the left panel is due to the contribution of the baryonic component.

This is confirmed by the results shown in the right-hand panel of Fig~\ref{Fig:vcirc_sgm_re}, which plots the stellar half-mass radius $r_{\rm half}$ as a function of $M_*$. As discussed in detail by \citet{Ferrero-2021}, spheroid-dominated systems are more compact than disk-dominated galaxies, which implies that, when comparing two galaxies with the same stellar mass, the one with higher $f_{\rm sph}$ has a smaller size. As a result, the baryonic component typically dominates over the dark matter in that region, shifting $V_{\rm circ, tot}(r_{\rm half})$ to higher values. For example, if we compare the three example galaxies shown in Fig.~\ref{Fig:gal_example}, the spheroid-dominated galaxy (open square) has a circular velocity $48\%$ higher than the disk-dominated one (open circle), mainly because the latter is roughly 5 times larger in terms of stellar half-mass radius.

\section{Properties of simulated disks and spheroids}
\label{Sec:Disk and spheroid properties}

We turn our attention now to the scaling relations of disks and spheroids, considered independently of the importance of the other component. In other words, we explore how the scaling relations discussed in the preceding section change if we only consider variations solely as a function of disk mass, or spheroid mass, respectively. 

\subsection{Mass vs characteristic velocity}

\begin{figure*}
    \centering
    \begin{subfigure}{0.497\textwidth}
        \centering
        \includegraphics[width=\textwidth]{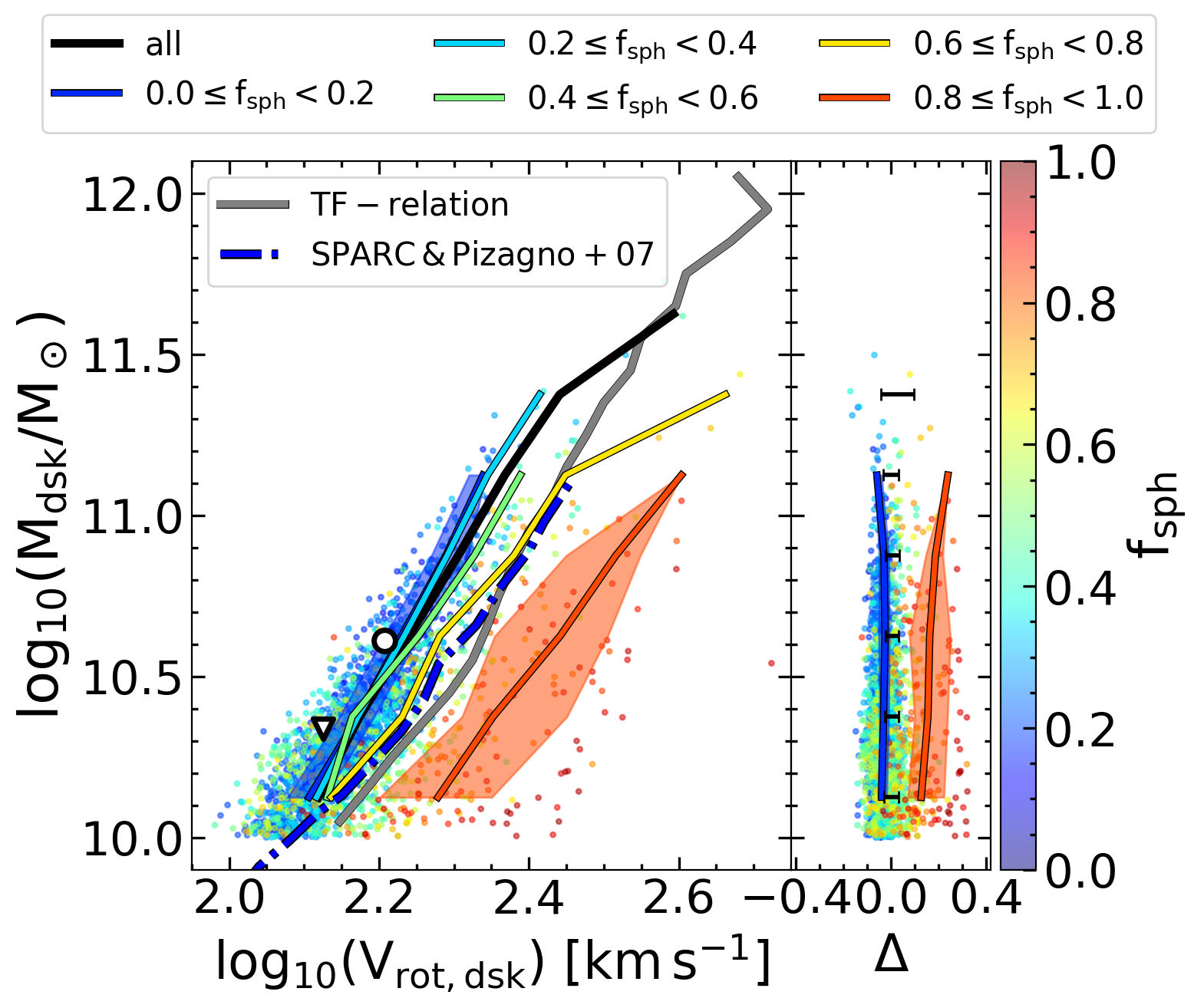}
    \end{subfigure}
    \hfill
    \begin{subfigure}{0.497\textwidth}
        \centering
        \includegraphics[width=\textwidth]{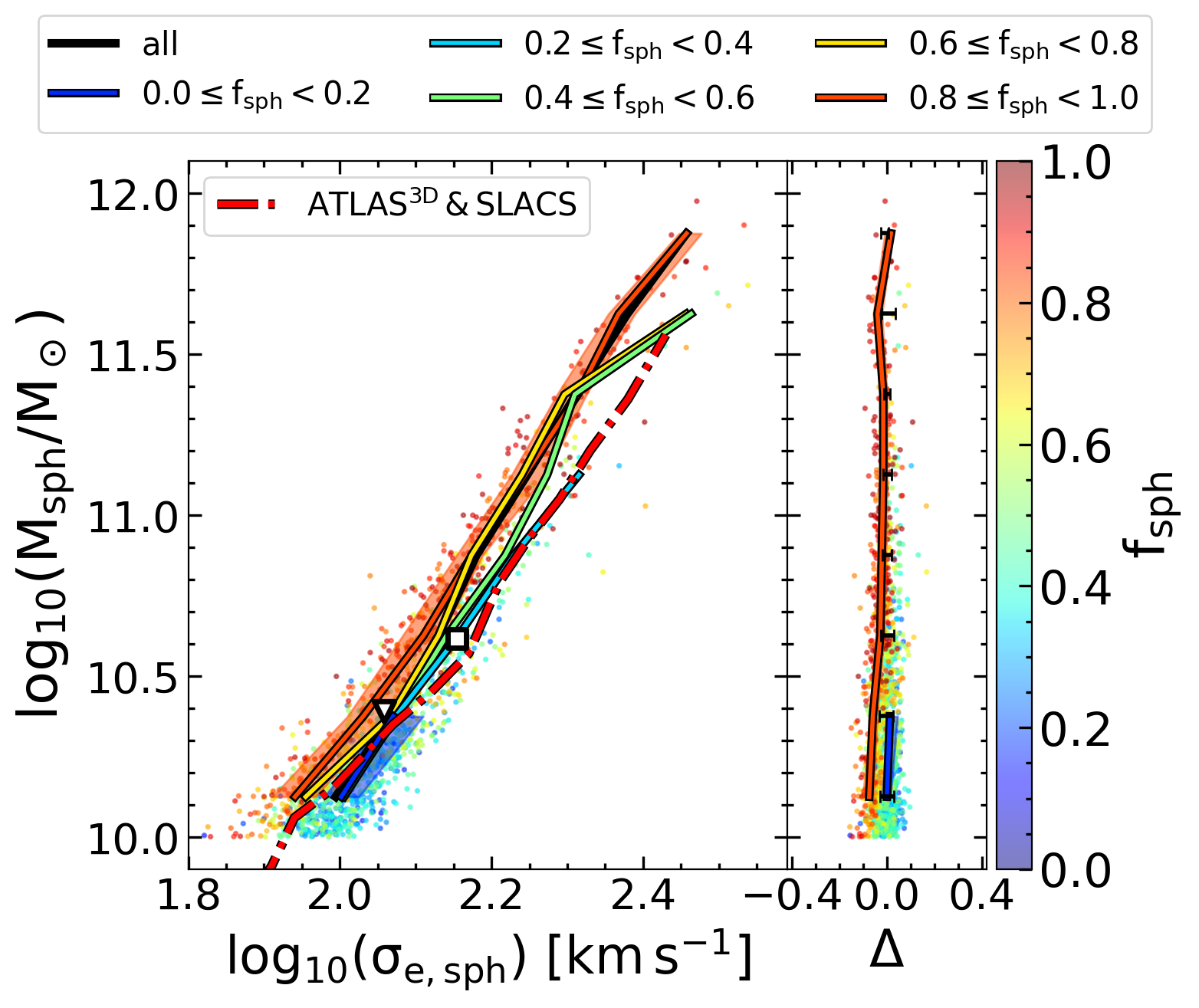}
    \end{subfigure}
    \caption{Scaling relations of disk and spheroidal components, treated separately. {\it Left panel}: Disk mass versus disk rotation velocity, colored by spheroidal mass fraction, as shown in the colorbar. The gray solid line corresponds to the TNG100 Tully-Fisher relation, shown in the left panel of Fig.~\ref{Fig:vcirc_sgm_re}, as a reference. The blue dot–dashed line shows the median TF-relation from \citet{Ferrero-2021} for a galaxy sample from the SPARC compilation \citep{Lelli-2016} and from the catalog of \citet{Pizagno-2007}. The right subpanel displays the ratios with the median values ($\Delta$) with error bars denoting the overall 25th and 75th percentiles of the data distribution, regardless of $f_{\rm sph}$. {\it Right panel}: Spheroid stellar mass versus spheroid velocity dispersion, colored by spheroidal mass fraction, as shown by the colorbar. The red dot–dashed line shows the median FJ-relation from \citet{Ferrero-2021} for an elliptical galaxy sample from $\mathrm{ATLAS^{3D}}$ \citep{Cappellari-2013a, Cappellari-2013b} and SLACS \citep{Bolton-2006}. The subpanel follow the same convention as the left panel. In both panels, the black line shows the median velocity trend of the entire sample, and colored solid lines show medians in bins of $f_\mathrm{sph}$ with shaded areas enclosing the 25th and 75th percentiles. The white symbols highlight the example galaxies shown in Fig.~\ref{Fig:gal_example}, with $f_\mathrm{sph}= 0.13,\ 0.53,\ 0.86$, respectively.}
    \label{Fig:scaling-relations-comp}
\end{figure*}

The Tully-Fisher \citep[TF;][]{Tully-1977} and Faber-Jackson \citep[FJ;][]{Faber-1976} relations encapsulate the dependence of the characteristic velocity of a galaxy on its stellar mass. How are these relations affected by the relative importance of a disk component (in the case of FJ) or of a spheroid (TF)?

The left panel of Fig.~\ref{Fig:scaling-relations-comp} shows the dependence of $M_{\rm dsk}$ on the disk mean rotation velocity at its half-mass radius, $V_{\rm rot, dsk}$). The latter is computed as the mean tangential velocity of all disk particles close to the disk mid-plane. Points in this panel correspond to individual disks, colored by the importance of its spheroidal component ($f_{\rm sph}$). Recall that, at fixed $M_{\rm dsk}$, increasing $f_{\rm sph}$ implies an increase in the total mass of the galaxy. 

We have also included in Fig.~\ref{Fig:scaling-relations-comp} the median of the observed TF-relation for disk-dominated galaxies (dot-dashed blue line) from \citet{Ferrero-2021}, based on a sample of galaxies from the SPARC compilation \citep{Lelli-2016} and from the catalog of \citet{Pizagno-2007}. At fixed stellar mass, the rotational velocity of both dynamically identified disks and simulated galaxies seems comparable with that inferred from observations.

The data on the left panel of Fig.~\ref{Fig:scaling-relations-comp} show that, at fixed $M_{\rm dsk}$, disk rotation velocities increase steadily as $f_{\rm sph}$ increases; in other words, the ``zero-point" of the relation shifts to larger velocities as $f_{\rm sph}$ increases. This is because, as we saw in the left panel of Fig.~\ref{Fig:vcirc_sgm_re}, the circular velocity increases systematically with $M_*$, and we expect the mean rotation velocity of the disk to trace approximately the circular velocity of the system at its half-mass radius. In other words, the characteristic velocity of disks is not a property of the disk component itself, but rather of the whole galaxy. The presence of the spheroid has a strong influence on the TF-relation of the disk component.

On the other hand, the Faber-Jackson relation of spheroids (see right-hand panel in Fig.~\ref{Fig:scaling-relations-comp}) is tight and well defined, and appears to be less affected by the presence of a disk. Indeed, at fixed $M_{\rm sph}$, the velocity dispersion of spheroid stars does not appear to show a strong dependence, on average, on $f_{\rm sph}$. This is somewhat surprising, given our results for the disk TF-relation, but perhaps not entirely unexpected. Spheroids are rather compact and, in $\sim$80\% of the cases in the sample, tend to dominate the gravitational potential inside their stellar half-mass radius. 

This may be seen in the middle panel of Fig.~\ref{Fig:vcirc_sgm_re}, where the thick gray line shows that the total circular velocity at $r_{\rm half}$ is roughly twice as large as the contribution of the dark matter alone in spheroid-dominated galaxies. Disks, on the other hand, are significantly more extended (see the right-hand panel of Fig.~\ref{Fig:vcirc_sgm_re}), which reduces their influence on their more compact central spheroids. 

These two facts imply that the addition of a disk has a limited effect on spheroid stars, whose kinematics seems to depend mainly on the mass and size of the spheroidal component alone.

Spheroid-dominated galaxies seem to follow a similar FJ-relation as TNG spheroids, as shown by the red dot-dashed line in the right-hand panel of Fig.~\ref{Fig:scaling-relations-comp}, which corresponds to the median of the observed FJ-relation taken from \citet{Ferrero-2021}, and based on an elliptical galaxy sample from $\mathrm{ATLAS^{3D}}$ \citep{Cappellari-2013a, Cappellari-2013b} and from the Sloan Lens ACS (SLACS) Survey \citep{Bolton-2006}. There seems to be a slight offset between observed and simulated FJ relations, in the sense that at a given velocity dispersion, observed galaxies are a bit more massive. The effect, however, is small, amounting to only $0.26$ dex in $M_{\rm sph}$.

\subsection{Mass vs size}
\label{Subsec:r_half}

\begin{figure}
    \centering
    \includegraphics[width=\columnwidth]{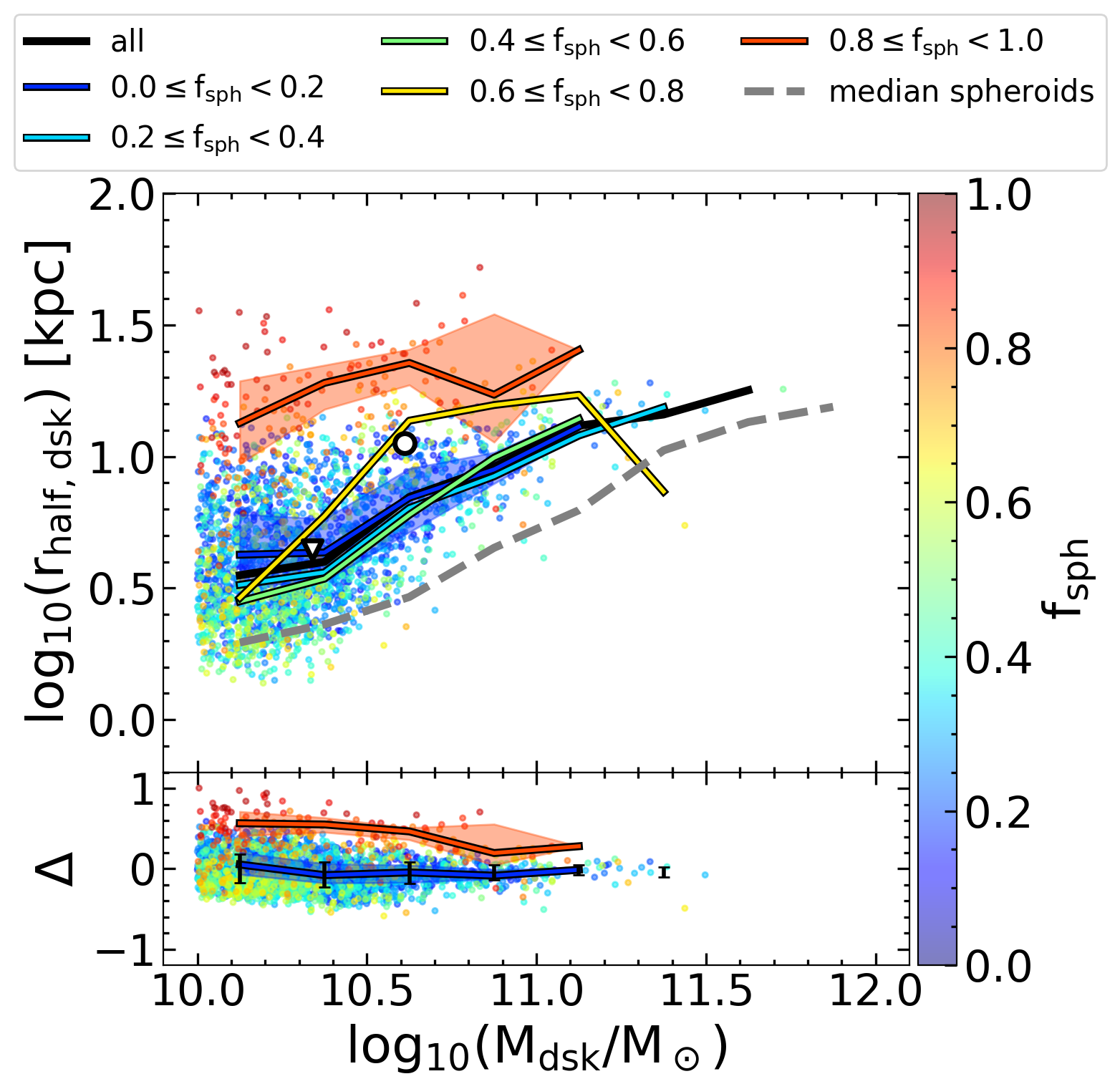}
    \includegraphics[width=\columnwidth]{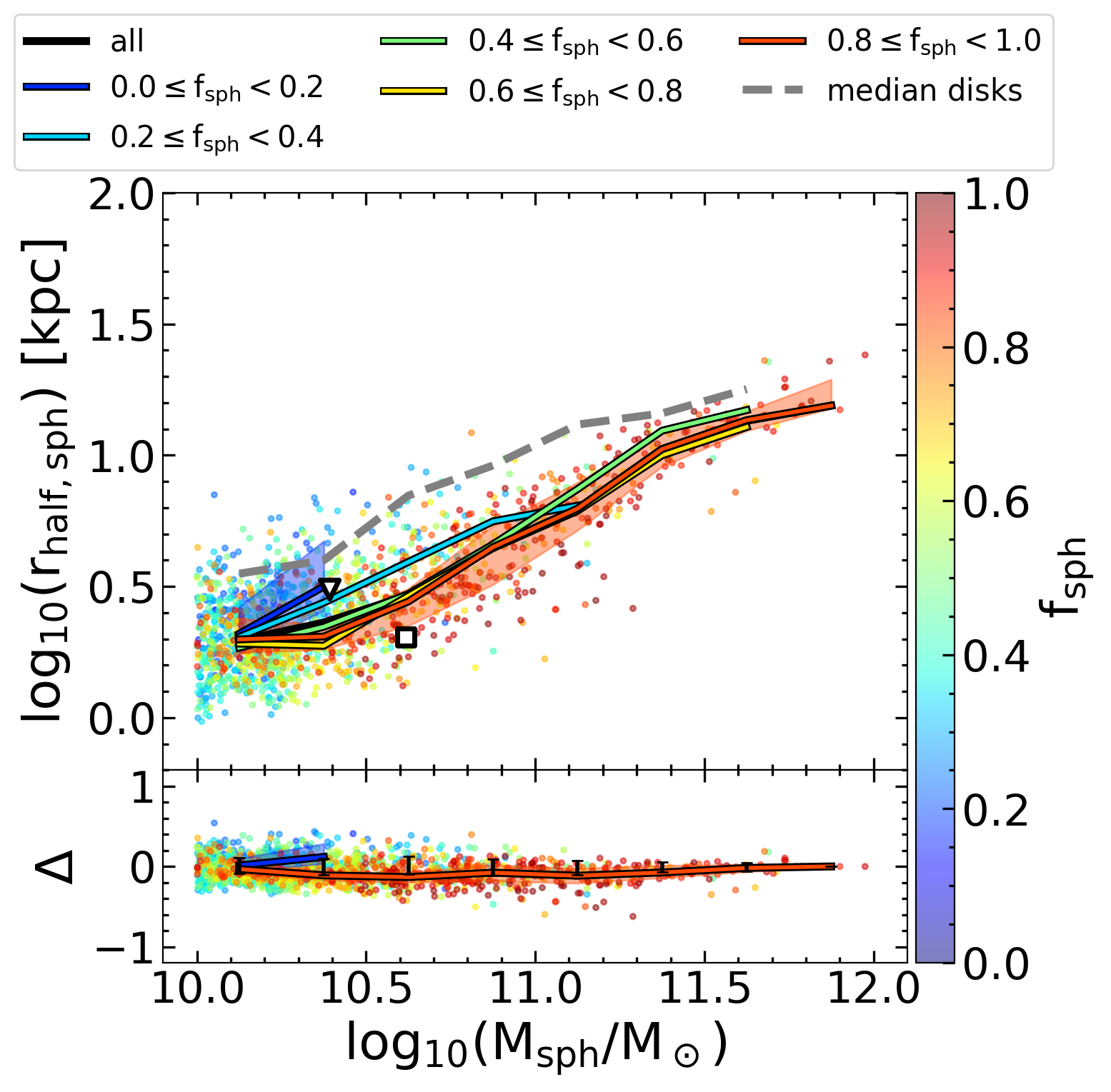}
    \caption{Stellar half-mass radius as a function of stellar mass for disks (top panel) and spheroids (bottom panel). {\it Top panel:} Disks, colored by spheroidal mass fraction $f_{\rm sph}$. Solid black line represent the median trend for the full sample calculated in bins of $\log_\mathrm{10}(M_\mathrm{dsk}/M_\odot)$, while colored solid lines show the median trends in bins of $f_{\rm sph}$, with shaded areas enclosing the 25th and 75th percentiles for selected bins. The gray dashed line indicates the median relation of the spheroidal component for reference. White symbols highlight the example galaxies shown in Fig.~\ref{Fig:gal_example} ($f_{\rm sph}= 0.13, 0.53, 0.86$). The bottom subpanel displays the ratios ($\Delta$) relative to the median values, with error bars denoting the 25th and 75th percentiles of the distribution. {\it Bottom panel}: Spheroids. Lines, shaded regions, color-coding, white symbols, and the subpanel follow the same convention as the top panel, with the gray dashed line showing the median relation of the disk component for comparison.}
    \label{Fig:Redsk_M200}
\end{figure}

Size is another key physical property of disks and spheroids. We analyze this in Fig.~\ref{Fig:Redsk_M200}, where the top and bottom panels show stellar half-mass radius as a function of mass for disks and spheroids, respectively. As in previous figures, we color results by the importance of the spheroid ($f_{\rm sph}$).

As in the case of the characteristic velocity, Fig.~\ref{Fig:Redsk_M200} shows that disk size is sensitive to the presence of a spheroidal component. Indeed, at fixed $M_{\rm dsk}$, $r_{\rm half, dsk}$ can be $3$-$4$ times higher in a spheroid-dominated galaxy ($f_{\rm sph}>0.8$) than in a disk-dominated system ($f_{\rm sph}<0.2$). In other words, the size of the disk component is largely set by the total mass of the galaxy, rather than by the disk component itself.

This is not the case for spheroids (bottom panel of Fig.~\ref{Fig:Redsk_M200}), whose size depends on spheroid mass and is less affected by the presence of a disk. Like its velocity dispersion, the size of spheroidal components is apparently set by the mass of the spheroid alone, rather than by the total stellar mass of the galaxy.

Another interesting result from Fig.~\ref{Fig:Redsk_M200} is that, at fixed stellar mass, disks are systematically larger than spheroids, as shown by the gray curves in each panel, which correspond to the median trend seen in the other panel. At $10^{10.5}\, M_\odot$ the median $r_{\rm half,dsk}$ is roughly $5$ kpc, or nearly twice as large as the median $r_{\rm half,sph}$ of spheroids of similar mass. This difference in size increases up to a factor of 8 when comparing the components of galaxies dominated by the spheroidal component ($f_{\rm sph}>0.8$ or, equivalently, $f_{\rm dsk}<0.2$).

\subsection{Formation times}

\begin{figure}
    \centering
    \includegraphics[width=\columnwidth]{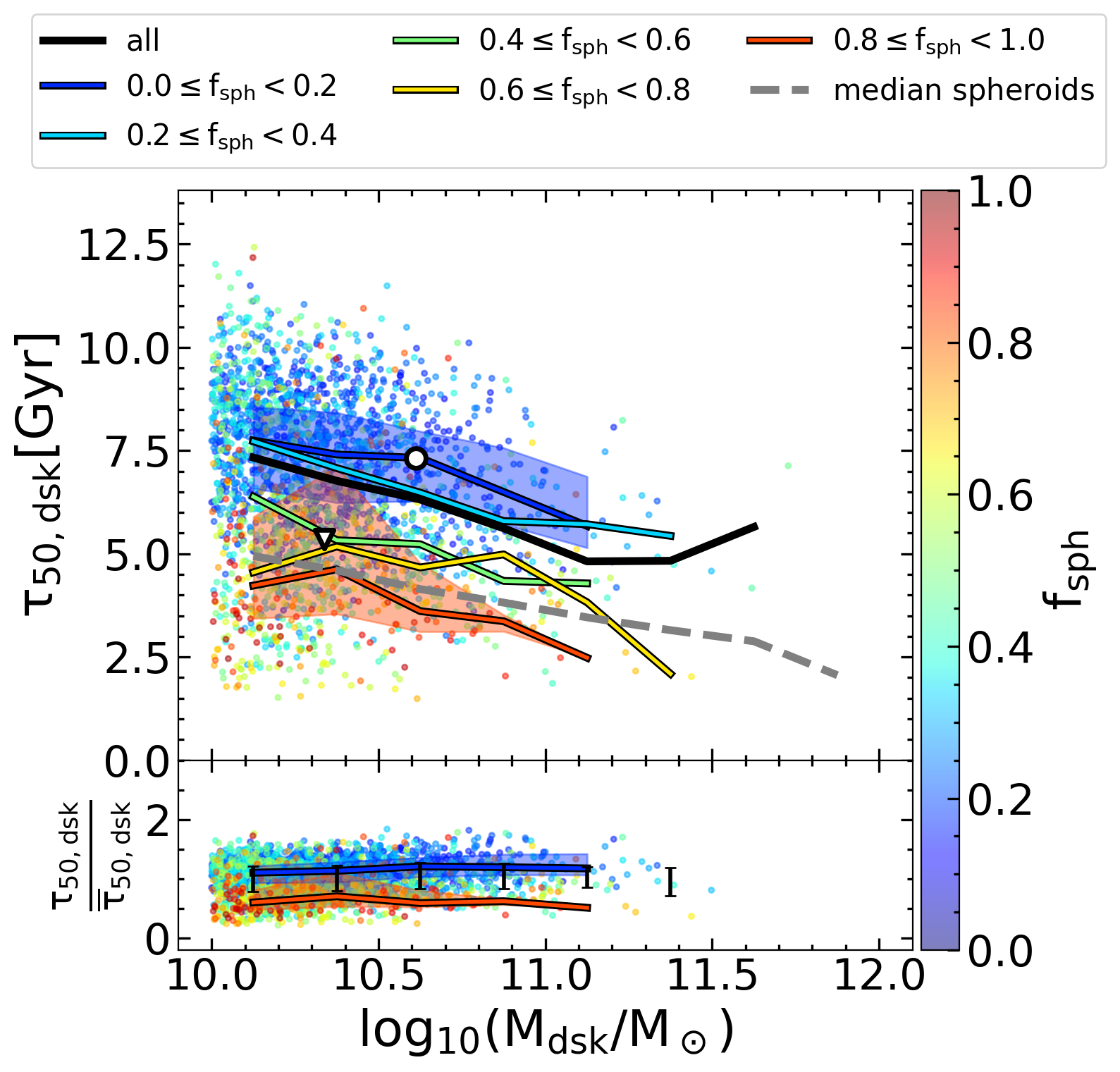}
    \includegraphics[width=\columnwidth]{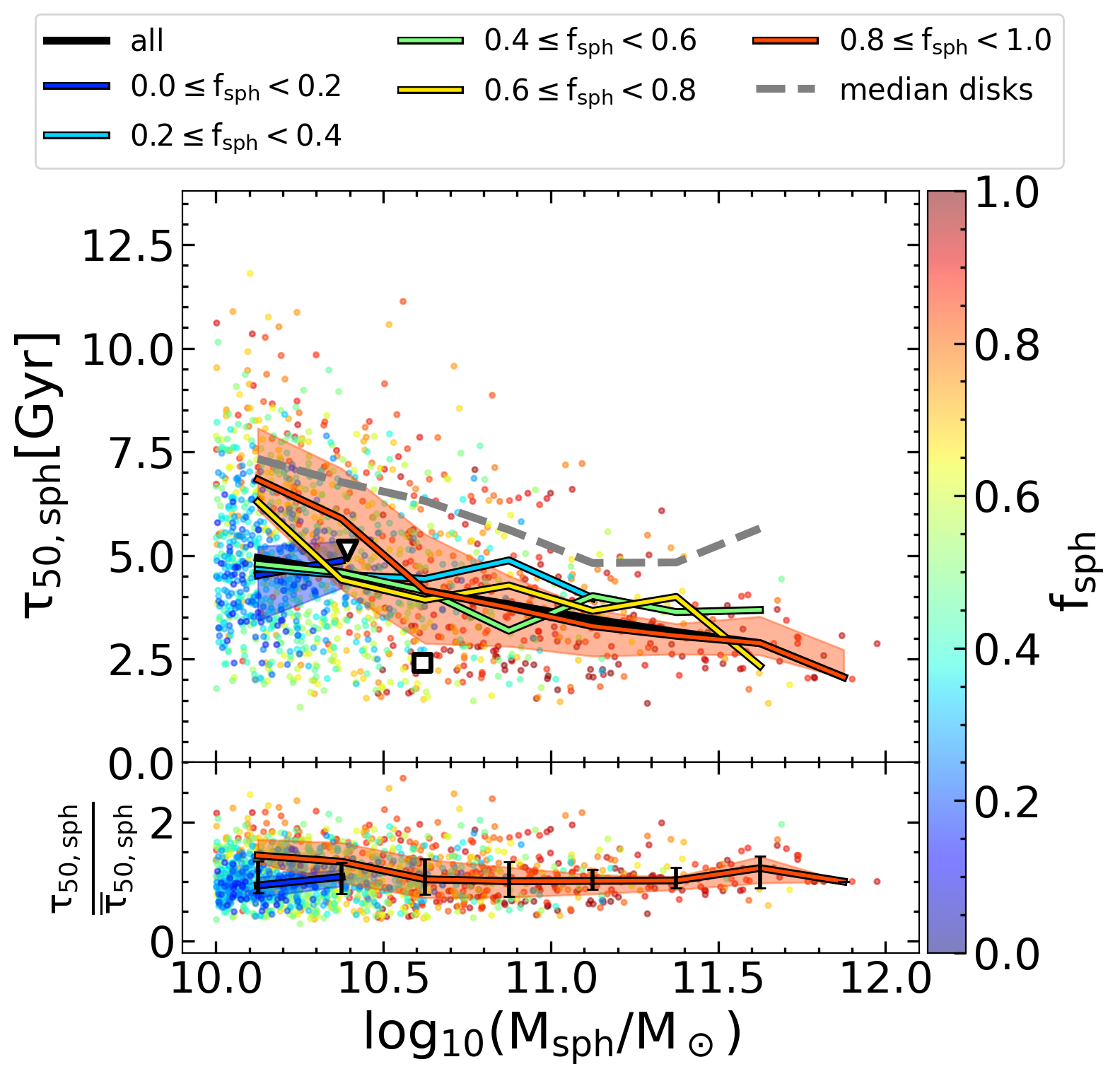}
    \caption{Half-mass formation time as a function of mass for disks (top panel) and spheroids (bottom panel). {\it Top panel:} Disks. Points are color-coded by the spheroid mass fraction ($f_{\rm sph}$). Solid black lines represent the median trend for the full sample calculated in bins of $\log_{10}(M_\mathrm{dsk}/M_\odot)$, while colored solid lines show the median trends in bins of $f_{\rm sph}$, with shaded regions enclosing the 25th and 75th percentile for selected bins. The gray dashed line indicates the median relation of the spheroidal component for reference. White symbols represents the example galaxies shown in Fig.~\ref{Fig:gal_example} ($f_{\rm sph}= 0.13, 0.53, 0.86$). The bottom subpanel displays the ratio with the median value ($\tau_{50, \rm dsk}/\bar{\tau}_{50, \rm dsk}$), with error bars denoting the 25th and 75th percentiles of the full distribution. {\it Bottom panel:} Spheroids. Lines, shaded regions, color-coding, white symbols, and bottom subpanel follow the same convention as the top panel, with the gray dashed line showing the median relation of the disk component for comparison.}
    \label{Fig:t50_med}
\end{figure}

Another physical property that reflects the evolutionary history of a stellar component is the age distribution of its stellar population. We use the parameter $\tau_\mathrm{50}$ as an estimate of the characteristic formation time, defined as the cosmic time at which 50\% of the present-day stellar mass of a particular component had formed. 

The top and bottom panels of Fig.~\ref{Fig:t50_med} show $\tau_\mathrm{50}$ vs mass, for stellar disks and spheroids, respectively. Disk formation times correlate with mass, with massive disks forming earlier than low-mass ones. These formation times also depend sensitively on the importance of the spheroidal component. At a given $M_{\rm dsk}$, formation times decrease (i.e., they become older) as the importance of the spheroidal component increases alongside the total stellar mass of the galaxy. 

The trend is, again, different for spheroids, as shown in the bottom panel of Fig.~\ref{Fig:t50_med}. Although a trend for more massive systems to form earlier is also present, spheroid formation times show little dependence on the importance of a disk. As for the size and characteristic velocity, the formation time of a spheroid seems to be set, on average, mainly by the mass of the spheroid itself. 

Another interesting result to be gleaned from Fig.~\ref{Fig:t50_med} is that, at given mass, spheroids form earlier than disks. This is highlighted by gray lines in each panel, which shows the median trend of the other panel in the same figure. For example, at $10^{10.5}\, M_\odot$, the disk median formation time is $\sim 6.5$ Gyr, compared with $\sim 4$ Gyr for a spheroid of the same mass. Only spheroid-dominated disks (i.e., $f_{\rm sph}>0.8$) form as early as spheroids.

\section{Conclusions}
\label{Sec:Conclusions}

We perform a dynamical decomposition of 3452 central galaxies from the TNG100 cosmological simulations into a disk stellar component and a spheroid stellar component. This procedure enables a separate analysis of the physical properties of the stellar disks and stellar spheroids components, as a function of the global galactic properties. 

We find that 54\% of the total stellar mass in galaxies with $M_*>10^{10}\, M_\odot$ reside in spheroids, while the remaining 46\% belong to disks. These percentages agree with observational estimates based on morphological classifications \citep{Kelvin-2014, Moffett-2016a} and photometric decompositions \citep{Schechter-1987, Moffett-2016}. The spheroid mass fraction increases with total stellar mass, from $f_\mathrm{sph}\sim 0.3$ at the low-mass end to $f_\mathrm{sph}\sim 0.9$ at the high-mass end, consistent with those obtained observationally by \citet{Moffett-2016}. This trend is reflected in the galaxy stellar mass function, where disk-dominated galaxies ($f_\mathrm{sph} < 0.5$) prevail at stellar masses below $\log_\mathrm{10}(M_*/M_\odot) \lesssim 11.0$, while spheroid-dominated galaxies ($f_\mathrm{sph} \geq 0.5$) dominate above this threshold.

At fixed halo mass, disk-dominated galaxies are more massive than spheroid-dominated galaxies, which could be interpreted as a result of disks forming stars over a longer period of time than spheroids, which get quenched earlier on. Equivalently, at fixed stellar mass, spheroid-dominated galaxies reside in halos that are roughly twice as massive as those that inhabit disk-dominated galaxies.

Regarding the physical properties of the disk and spheroid populations dynamically identified, our results may be summarized as follows:
\begin{itemize}
    \item At fixed $M_{\rm dsk}$, the rotation speed of the disk increases systematically with $f_{\rm sph}$, which implies that the ``Tully-Fisher relation" of disks depends mainly on the total mass of the galaxy rather than that of the disk alone. In contrast, at fixed $M_{\rm sph}$, the stellar velocity dispersion depends little on $f_{\rm sph}$, which suggests that the ``Faber-Jackson" relation of spheroids depends almost exclusively on the mass of the spheroid, and not on the total stellar mass of the galaxy.

    \item As in the previous item, the size of simulated disks is found to be quite sensitive to the total mass of a galaxy, whereas the size of a spheroid seems to depend mainly on the spheroid mass rather than the total stellar mass.

    \item A similar result is also found for the formation time of each component. At  fixed $M_{\rm dsk}$, disks get older as $f_{\rm sph}$ increases, again indicating that the properties of the disk are sensitive to the relative importance of the spheroid and, as a result, to the total stellar mass of the galaxy. Spheroid formation times, on the other hand, are largely set by the mass of the spheroid, rather than the total stellar mass of the galaxy.
     
\end{itemize}

In summary, disk properties are sensitive to the importance of the spheroid with which they coexist. In contrast, the properties of spheroids depend almost entirely on their own stellar mass, a result that we associate with the fact that spheroids are systematically more concentrated than disks. To first order, then, the bulge of a simulated disk galaxy shares broad similarities with a spheroidal galaxy of the same stellar mass, an intriguing result whose proper interpretation we defer to future work.

\begin{acknowledgements}
    We would like to thank the anonymous reviewer for their comments and suggestions that improved the quality of the paper. VAC and MGA thank the financial support from Agencia Nacional de Promoción Científica y Tecnólogica, the Consejo Nacional de Investigaciones Científicas y Técnicas (CONICET, Argentina) and the Secretaría de Ciencia y Tecnología de la Universidad Nacional de Córdoba (SeCyT-UNC, Argentina). This project has received funding from the European Union’s HORIZON-MSCA2021-SE-01 Research and Innovation programme under the Marie Sklodowska-Curie grant agreement number 101086388 – Project acronym: LACEGAL. VAC and JFN wish to thank the Donostia International Physics Center (DIPC), and especially to Silvia Bonoli and Raúl Angulo for their support. JFN acknowledges support from a Royal Society Wolfson Visiting Fellowship at Durham University and from the Humboldt Foundation at the Max-Planck Institute for Astrophysics.
\end{acknowledgements}

\bibliographystyle{aa}
\bibliography{bibliography}

\end{document}